# Multi-modal Mining and Modeling of Big Mobile Networks Based on Users Behavior and Interest

Saeed Moghaddam, Ahmed Helmy

Computer and Information Science and Engineering Department
University of Florida
moghaddam@ufl.edu, helmy@ufl.edu

***Abstract*— Usage of mobile wireless Internet has grown very fast in recent years. This radical change in availability of Internet has led to communication of big amount of data over mobile networks and consequently new challenges and opportunities for modeling of mobile Internet characteristics. While the traditional approach toward network modeling suggests finding a generic traffic model for the whole network, in this paper, we show that this approach does not capture all the dynamics of big mobile networks and does not provide enough accuracy. Our case study based on a big dataset including billions of netflow records collected from a campus-wide wireless mobile network shows that user interests acquired based on accessed domains and visited locations as well as user behavioral groups have a significant impact on traffic characteristics of big mobile networks. For this purpose, we utilize a novel graph-based approach based on KS-test as well as a novel co-clustering technique. Our study shows that interest-based modeling of big mobile networks can significantly improve the accuracy and reduce the KS distance by factor of 5 comparing to the generic approach.**

***Keywords- user interest; mobile data; traffic; big data; co-clustering***

## I. Introduction

Mobile Internet traffic has experienced a significant growth in the recent years. Different types of Internet-enabled mobile devices are getting more and more popularity and wireless Internet access infrastructures are growing faster than ever. The emergence of this radical change in availability of Internet raises a new need for modeling of Internet characteristics in big mobile networks. A traffic model in general is a model that can be used to regenerate the behavior of a real traffic stream. A major application of traffic models is in predicting the behavior of traffic as it passes through a network. The common approach toward traffic modeling is to find a generic model for the whole network. Although, such models provide good approximations for the old wired Internet, but several studies have shown that they do not fit the dynamics of wireless networks. For example, [1] characterizes the wireless traffic in different locations and shows that the dynamics of network follow a similar model but with different parameters. However, such models are generally based on small datasets of WLAN activities (e.g. 25000 flows a day), which are far from the full scale of dynamics in current big mobile networks (e.g., our dataset includes over 100 million flows per day). Moreover, most previous works have not studied the characteristics of big mobile networks based on user behavior and interests which can be acquired based on accessed domains (e.g. ‘cnn’) or visited locations (e.g. ‘cinema’). Interest-based and behavior-aware modeling of big mobile network traffic can be beneficial to the realistic design of applications, protocols and services (e.g. for resource allocating or content caching).

In this paper, we present a novel modeling approach based on our earlier work on graph-based traffic analysis of domains and locations [2] and also introduce a novel technique for analysis and modeling of multi-modal user behavioral groups using a co-clustering approach. Our campus-wide case study shows that domains, locations and users have specific traffic characteristic that can also form groups with distinct characteristics. In our study, we investigate interest-based characteristics by partitioning the Internet traffic based on domains, buildings and user behavioral groups and analyzing the traffic characteristics using KS (Kolmogorov-Smirnov) test [3].

This work has the following key contributions:

1. We provide a novel interest-based traffic modeling technique for big mobile networks based on accessed domains (top 100 active domains) and visited locations (68 different buildings) across a campus with more than 32000 users. The studied dataset is one of the largest wireless mobile network traffic traces (including around 100 million records per day).

2. We provide a systematic method to discover similarities and differences between the traffic distributions of different domains or locations. We show how a novel graph-based technique can be applied to identify groups of domains or locations with distinct traffic characteristics.

3. We provide a novel technique to discover multi-modal user behavioral groups based on both domains and locations visitations. We show how a co-clustering technique based on information theory can be utilized to identify behavioral groups with distinct traffic characteristics.

4. We show that the proposed interest-based approach can significantly improve the accuracy of traffic modeling for big mobile network and reduce the KS distance by factor of 5 using KS-test and weighted traffic intensity of domains and locations.

The rest of the paper is organized as follows. In Section 2, we review the related work. In Section 3, we briefly describe the datasets and big data processing details. Section 4 presents our interest-based and behavioral modeling approach. Section 5 discusses the accuracy of the proposed approach and Section 6 concludes the paper.

## II. RELATED WORK

There has been many works on Internet traffic modeling and among all flow-level modeling has been one of the most popular approaches [4,5]. However, most of such studies use idealized models, e.g., Poisson process, to characterize flows. While such simplified models may be fine for the old wired Internet, they are not appropriate for big mobile wireless networks. Among the works looking into heavy-tailed distributions, [6] propose to use several heavy-tailed distribution models to characterize the statistical process associated with TCP flows in a wide-area network. In [7] Feldmann suggests Weibull distribution as a better fit than other distribution for modeling of wired TCP flow arrivals. However, these works are mainly based on small wired network and few studies have focused on traffic modeling for big mobile networks. While several works have characterized user and mobility patterns in wireless networks, most of them focused on host-level rather than flow-level. In [8, 9] Tang et al. studied users, network activity and host mobility patterns in a metropolitan-area wireless network and also on a campus department. Other studies at [10, 11] investigated wireless user and AP/building activity and aggregate traffic for the Dartmouth campus wireless network. In [12] Balazinska et al. studied user population characteristics, network usage and load distribution in corporate networks and in [13] Balachandran et al. characterized the aggregate network load and utilization and user patterns during a conference. For flow-level modeling of wireless networks, [1] propose a Weibull regression model to approximate the flow arrivals at individual APs. In another work, [14] found that accessed information by HTTP queries shows spatial locality in a wireless campus network.

On behavioral analysis of mobile networks, there has been a widespread interest for understanding the user behavior. The scope of analysis includes WLAN usage and its evolution across time [9-11] and user mobility [12,15]. Some works focus on using the observed user behavior characteristics to design realistic and practical mobility models [16,17]. In [1] it was shown that the performance of resource scheduling and TCP vary widely between trace-driven analysis and non-trace-driven model analysis. Several other works focus on classifying users based on their mobility periodicity [18], time-location information [19], or a combination of mobility statistics [8]. The work on the TVC model [16] provides a data-driven mobility model for protocol and service performance analysis. In other works, different techniques have been proposed for multi-dimensional modeling of users' interest and behavior based on co-clustering [20], self-organizing maps [21-23], Gaussian mixture model [24], domain/location specific modeling [25], global/local modeling [26], and frequent pattern mining [27, 28]. The key difference between the previous studies and this work is to provide an interest-based flow-level traffic modeling approach based on accessed domains and visited locations for big mobile networks.

The two main trace libraries for the networking communities can be found in the archives at [29] and [30]. None of the available traces provides big netflow data coupled with DHCP and WLAN sessions to be able to map IP addresses to MAC addresses and to AP, building and eventually to a context (e.g., history department or a fraternity). Our dataset is significantly larger and richer in semantic than the other mobile wireless network traces and includes around 100 million records per day. Our novel data-driven approach can develop realistic interest-based traffic models to enhance the performance of networking services design for big mobile networks.

One network application for interest-based traffic modeling is profile-based services. Profile-cast [31] provides a new one-to-many communication paradigm targeted at a behavioral groups. In the profile-cast paradigm, profile-aware messages are sent to those who match a behavioral profile. Behavioral profiles in [31] use location visitation preference and are not aware of Internet activity and traffic. However, iCast [32] provides an interest-aware implicit multicast approach for opportunistic mobile data dissemination based on online activities. In this work, we provide interest-based and behavior-aware traffic models that can be utilized to improve the design and evaluation of interest-based and behavior-aware services and protocols.

## III. BIG NETWORK DATA PROCESSING

Realistic traffic modeling and analysis of big mobile networks requires processing of big amount of network traces. In our study, we process extensive traces collected via all network switches around the campus including netflows, DHCP and wireless session logs. A flow is defined as a unidirectional sequence of packets with some common properties (e.g., source IP address) that pass through a network device (e.g., router) which can be used for flow collection. Network flow records include the start and finish timestamps, source and destination IP addresses, port numbers, protocol numbers, and flow sizes. The source and destination IP addresses combined with DHCP logs can be used to identify user device MAC addresses and the websites accessed respectively. The DHCP log contains the dynamic IP assignments to MAC addresses and includes date and time of each event. The wireless session log collected by each wireless access point (AP) includes the 'start' and 'end' events for device associations which can be used to derive users' location.

The variety and scale of different described traces is a major processing challenge (our dataset includes around 100 million flow records per day). To resolve this problem, we leveraged DataPath [33], a big data processing engine developed at University of Florida and Rice University. DataPath allows complex queries to be defined and executed over TB-sized data using novel techniques such as on-the-fly code generation, aggressive I/O, push-based data processing, hybrid column/row store and multi-threaded database operators. Furthermore, DataPath allows seamless integration of aggregation and mining tasks.

In our study, we first filtered the popular IP prefixes (first 24 bits) using a threshold (the reason for using 24 bits filter is the fact that popular websites usually use an IP range instead of a single IP address). Then, for the filtered IP prefixes, their domains were resolved. Among the resolvable domains, the top 100 active ones were identified and all the users interacting with those domains (e.g., 'google', 'facebook', etc.) were considered for the modeling phase. Then, the location of each Internet access (per flow) was identified using the WLAN session logs.

## IV. Interest-based and Behavioral Modeling

### A. Domain and Location Specific Analysis

In this section, we study the traffic behavior of big mobile wireless networks considering specific user interests in terms of accessed domains and visited locations. The goal of this study is to find similarities or differences between the behavior of mobile Internet traffic for individual domains or locations, and the overall traffic of the mobile wireless network. For this purpose, we first extract the flow-level traffic distribution for different domains and buildings (per second). Then, we examine the dataset against different statistical distributions to find the best curve fitted to the real distributions. The set of distributions includes Weibull, Rayleigh, Poisson, Negative Binomial, Lognormal, Generalized Pareto, Generalized Extreme Value, Exponential and Gamma. We pick the best fit based on the KS (Kolmogorov-Smirnov) test [3]. The KS test is a nonparametric test for the quality of continuous, one-dimensional probability distributions that can be used to compare a sample with a reference probability distribution. The KS statistic quantifies a distance between the empirical distribution function of the sample and the cumulative distribution function of the reference distribution. The KS-test has the advantage of making no assumption about the distribution of data. In our experiment, we used a confidence level of 5 percent for the KS test. In addition to domain and location specific modeling, we also find the best fit for the overall traffic.

Our study shows that traffic behaviors of different domains follow different types of distributions which form four categories. On average, 25 percent of domains follow Weibull, 23 percent follow Lognormal, 21 percent follow Generalized Extreme Value and the rest follow other types of distribution. Our study also shows that traffic characteristics of different location are not the same as well. We can again find four major categories of buildings. On average, 35 percent of buildings follow Weibull, 25 percent follow Lognormal, 18 percent follow Generalized Extreme Value and the rest follow other type of distributions. This clearly shows that the best generic fit which is Generalized Extreme Value is not always the best model when considering specific domains or locations.

### B. Graph-based Analysis

In the section, we investigate the similarities and differences between the traffic distributions of different domains and locations. While some domains or locations might follow the same type of statistical distribution, their models might follow different parameters. On the other hand, finding the best fit for different domains or locations does not provide us with a quantitative measure to compare their traffic similarities. Therefore, in this part of our study, we provide a method to compare the actual traffic distributions of different domains or locations. For this purpose, we apply another flavor of KS-test that is called Two-sample KS test. The two-sample KS test is one of the most useful and general nonparametric methods for comparing two samples, as it is sensitive to differences in both location and shape of the empirical cumulative distribution functions of the two samples. This test compares the distributions of the values in the two input data samples. The null hypothesis is that the two samples are from the same distribution. The alternative hypothesis is that they are from different distributions. The two-sample Kolmogorov–Smirnov statistic for samples of size $n$ and $n'$ is:

$$D_{n,n'} = sup_x \left| F_{1,n}(x) - F_{2,n'}(x) \right| \quad (1)$$

where $sup_x$ is the supremum of the set of distances, $F_{1,n}$ and $F_{2,n'}$ are the empirical distribution functions of the first and the second sample respectively. The null hypothesis is rejected at significance level $\alpha$ if:

$$\sqrt{\frac{nn'}{n+n'}} D_{n,n'} > K_{\propto} \quad (2)$$

In our study, we run the test at significant level of 5 percent for each pair of domains or buildings. The results form two matrices including a 100*100 matrix for domains and another 68*68 for buildings showing if two domains or buildings follow the same distribution or not. In order to analyze the result, each of the matrices can be interpreted and visualized as traffic similarity graphs. In such graphs nodes represents domains or buildings with an edge between nodes if the corresponding domains or buildings have similar traffic distributions. Fig. 1 and Fig. 2 shows the resulting graph for domains and locations after running the algorithm presented in [34] for finding the modularity classes within the graphs and applying Fruchterman-Reingold algorithm [35] to form the graph layouts. In these graphs, the corresponding domain or building for each of the nodes can be found from the node identifier number (mappings between identifier numbers and domains and buildings are available in Fig. 3). Modules (or groups) are shown using different colors in the figures. The size of each node represents its degree in the graph that shows uniqueness of traffic distribution of the node compared to the other nodes (low degree is interpreted as uniqueness).

*1) Domain-based Analysis*

As can be seen in Fig. 1 for domain-based analysis, there are 21 domains with unique traffic characteristics. As can be observed, most of very popular domains including ‘google’, ‘facebook’ and ‘apple’ have unique characteristic. In other words, high traffic domains show more uniqueness in terms of their traffic characteristics.

The rest of domains form 12 groups with distinct traffic distributions. Half of the groups have a size of less than 5 and the size of rest is up to 16. Studying different groups reveals many interesting facts. For example, video sharing domains like ‘netflix’ and ‘veoh’ show unique characteristics. We can also observe traffic distributions of ‘cnn’, ‘msnbcsport’ and ‘microsoft’ (the group at top-left) are similar. The interesting fact here is that both ‘cnn’ and ‘msnbc’ provide news and on the other hand both ‘microsoft’ and ‘msnbc’ are provided by the same entity, i.e., Microsoft. This shows that the type of provided content by a domain and also its content provider may affect its traffic distribution. This might also show that, in some cases, when various types of domain attributes (e.g. content type and provider) are appeared together (as in ‘msnbcsport’) the traffic characteristics of the result is a combination of characteristics regarding those attributes.

Another example of interesting finding is that many of domains related to high-speed Internet and phone providers like ‘comcast’, ‘charter’ and ‘qwest’ have similar traffic distributions (the group in middle-right). Interestingly, we can also find ‘shoutcast’ in this group which is not in that category but provides a similar type of service, i.e., Internet radio stations (similar in the sense that both phone and radio services provide voice data). Our study shows that the traffic distribution of all domains in this group follows Rayleigh distribution which is different from the generic distribution.

*2) Location-based Analysis*

Fig. 2 shows the resulting graph for the location-based analysis showing three major groups of buildings with distinct traffic characteristics. By looking at the building categories we can again discover different interesting facts. For example, we can observe that more that 70 percent of buildings in Music, Cinema and Auditorium categories are in the same group (the group of nodes in right-bottom). We can also see most fraternities (9 ones) are in the same group with similar traffic characteristics (the big one at the left). This shows that type of location and its context has also have an important effect on characteristics of its traffic distribution. In other words, locations with similar context mostly follow similar traffic characteristics.

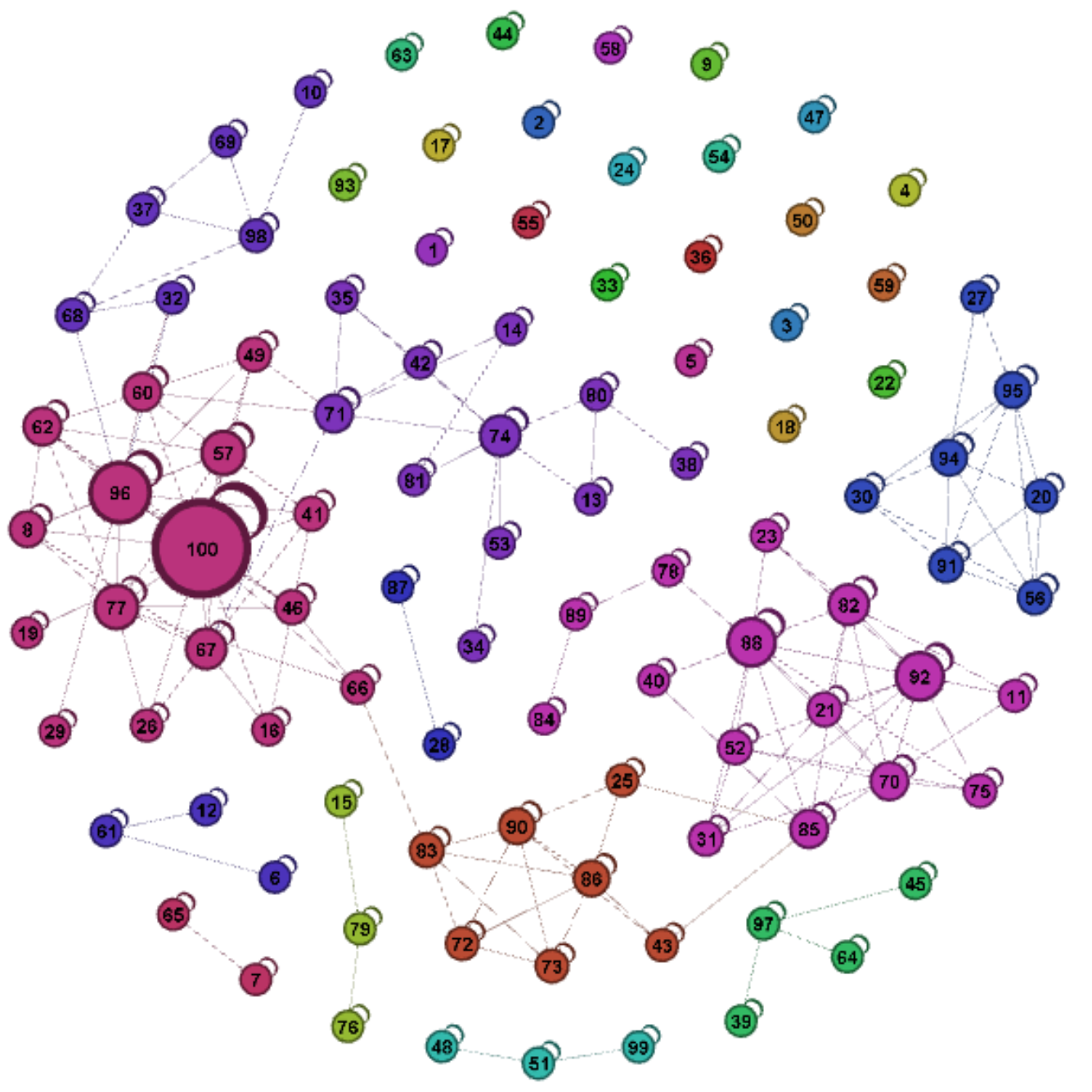

Fig. 1. Traffic similarity graph for domains. Nodes represent domains and show their IDs (Domain names can be found in Fig. 3). Colors show different detected modules (groups) within the graph. Size of each node shows its degree in the graph.

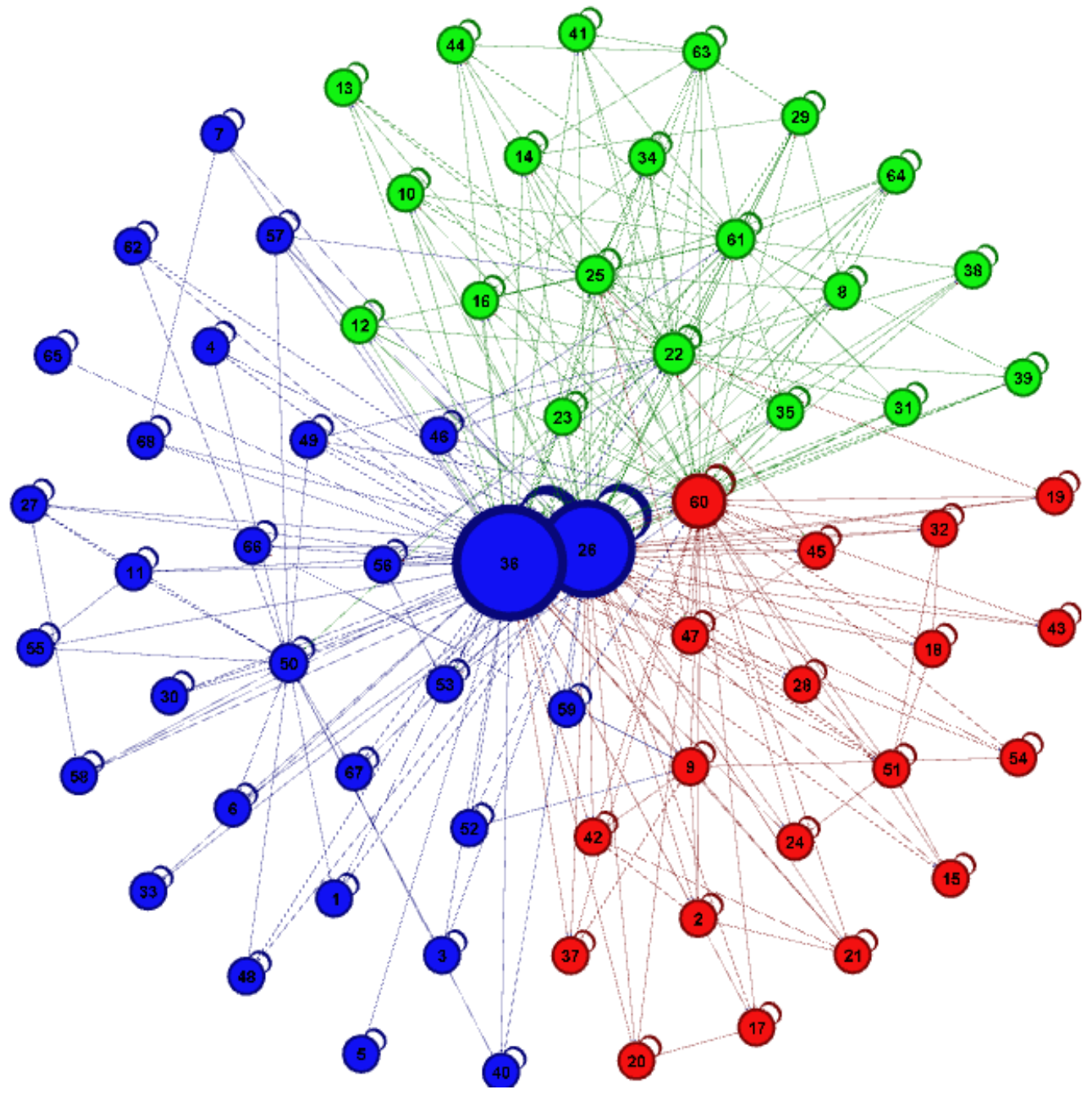

Fig. 2. Traffic similarity graph for locations. Nodes represent buildings and show their IDs (Building types can be found in Fig. 3). Colors show different detected modules (groups) within the graph. Size of each node shows its degree in the graph.

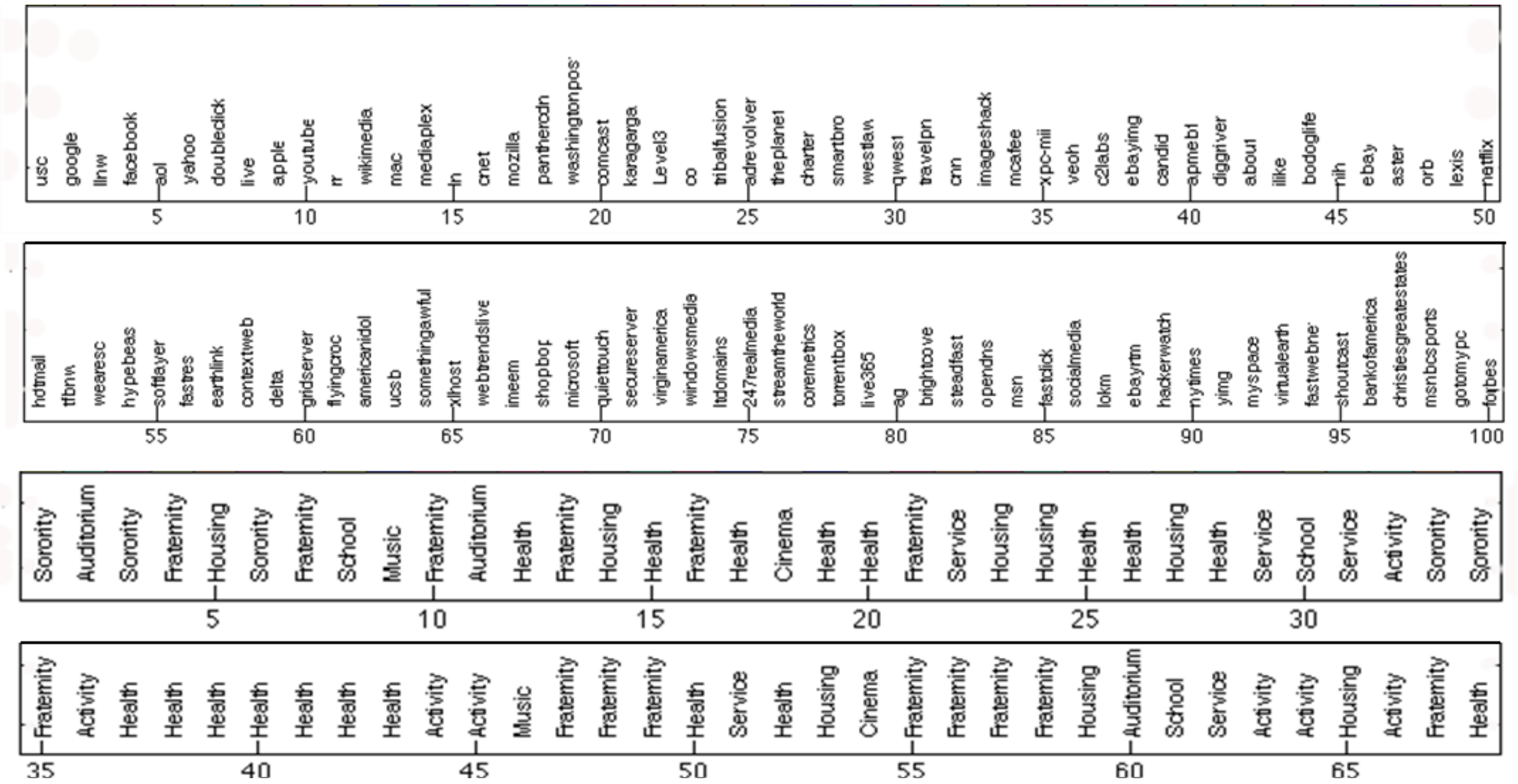

Fig. 3. Identifier mappings for domains and buildings

### C. *Multimodal Behavior-based Analysis*

In previous sections, we showed that different groups of domains and locations show distinct traffic characteristics. In this section, we study the traffic characteristics of user behavioral groups while considering traffic characteristics of different groups of domains or locations as well. By behavioral groups we mean groups of mobile users who show similar behavior in terms of domain access or location visitation patterns.

For this purpose, we first need to discover groups of users as well as groups of domains and locations with similar traffic characteristics and then investigate the characteristics of each group of users within each group of domains or locations. In the previous section, we used a graph-based technique to find groups of domains and locations with similar traffics; however, this approach is not scalable for thousands of users. Moreover, even if we find groups of users based on their overall traffic characteristics, it is not clear how we can correlate those groups with groups of domains or locations (an overall user group might not show similar characteristics when considering a specific group of domains or locations). To resolve these problems, we utilize a novel technique based on information-theoretic co-clustering [20] to discover multi-modal behavioral groups inside the mobile community. In the rest of this section, we first provide details of our novel technique and then discuss the traffic characteristics of multi-modal behavioral groups.

#### 1) *Mining of Behavioral Groups*

A very well-known approach to find groups of entities with similar characteristics (e.g., users, domains, locations) is to utilize clustering methods. However, a major challenge in finding multi-modal behavioral groups is the fact that ordinary one sided clustering algorithms like hierarchical clustering or k-means can only cluster data along one modality, e.g., we either get clusters of domains or clusters of users but not both at the same time. However, in our case, we need to find clusters of users and domains or locations concurrently so that we get a unified view on dynamics of traffic behavior across different multi-modal behavioral groups. For this purpose, we use the information-theoretic co-clustering technique [20] which clusters the input dataset along multiple modalities simultaneously. In this way, we can correlate different modalities in a unique model and identify distinct clusters of users-domains as well as users-locations.

The input data for the co-clustering algorithm is users' traffic data which represents flow-level traffic exchanged between users and different domains or different locations. Information-theoretic co-clustering technique treats the input data table as a joint-probability distribution of two discrete random variables, whose values are given in the rows and columns, and poses the co-clustering problem as an optimization problem in information theory. This technique defines mappings from rows to row-clusters and from columns to column-clusters and then tries to optimize the co-clustering result. The optimal co-clustering is one that leads to maximum mutual information between the clustered random variables, and minimizes the loss in mutual information between the original random variables and the mutual information between the clustered random variables.

This algorithm monotonically increases the preserved mutual information and optimizes the loss function. This task is performed by intertwining both row and column clustering. Column clustering is performed by calculating closeness of each column distribution (in relative entropy) to column cluster prototypes. Row clustering is performed similarly. This iterative process converges to a local minimum.

The algorithm never increases the loss, and so, the quality of co-clustering improves gradually. Iteratively, the method performs an adaptive dimensionality reduction and estimates fewer parameters than one-dimensional clustering approaches, resulting in a regularized clustering. In addition, the algorithm is efficient. The computational complexity of the algorithm is given by $O(N \bullet \tau \bullet (k + l))$ where k and l are the desired number of row and column clusters, N is the number of non-zeros in the input joint distribution and $\tau$ is the number of iterations.

In our study, we applied the co-clustering technique to find multi-modal behavioral groups based on both users-domains and users-locations traffics. The number of clusters can be set as the input parameters of the algorithm. In our case study, we tried different numbers of clusters and finally chose to form 10 clusters of users, domains or locations as it showed more distinct characteristics.

### 2) *Multi-modal Behavioral Analysis*

After determining the multi-modal behavioral groups, in the next step we find the best fit for the traffic distribution of each group based on the KS-test. Figure 4 shows the best fit for all the discovered behavioral groups in terms of domains and locations visitations. In the figure, each row represents a user group and each column represent a domain or location group. As can be observed, for user-domain groups, on average, Log-normal, Weibull and General Extreme Value are the best fits for 27, 25 and 24 percent of behavioral groups respectively. For user-location case, on average, Log-normal, Weibull and General Extreme Value are the best fits for 26, 26 and 20 percent of the none-empty groups respectively.

An interesting observation in figure 4-a is that the best traffic model for user-domain groups is majorly driven by domains not the users (several columns have the same best fit for most user groups). For example for domain groups #4, #6, and #9 more than 70 percent and for #3, #5, #7, #10 more than 50 percent of user groups follow same type distributions. However, in Fig 4-b, we cannot see a major dominance by the locations over users to drive the traffic characteristics. This shows that users have a relatively strong impact on the characteristic of multi-modal groups when considering location visitations.

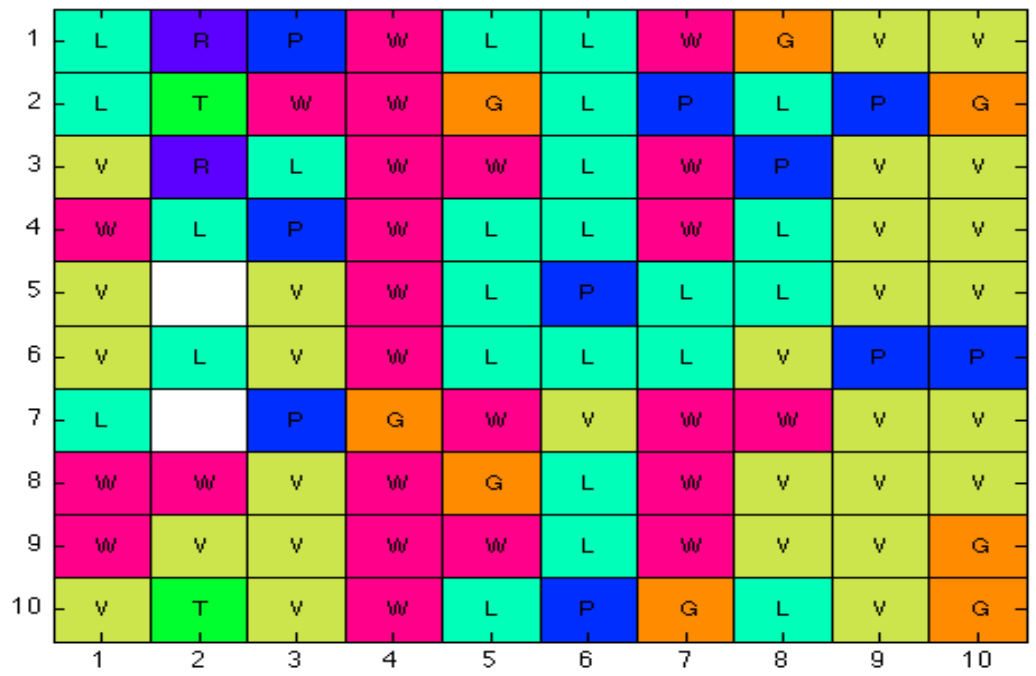

User-domain behavioral groups. X and y axes represent domain and user clusters respectively

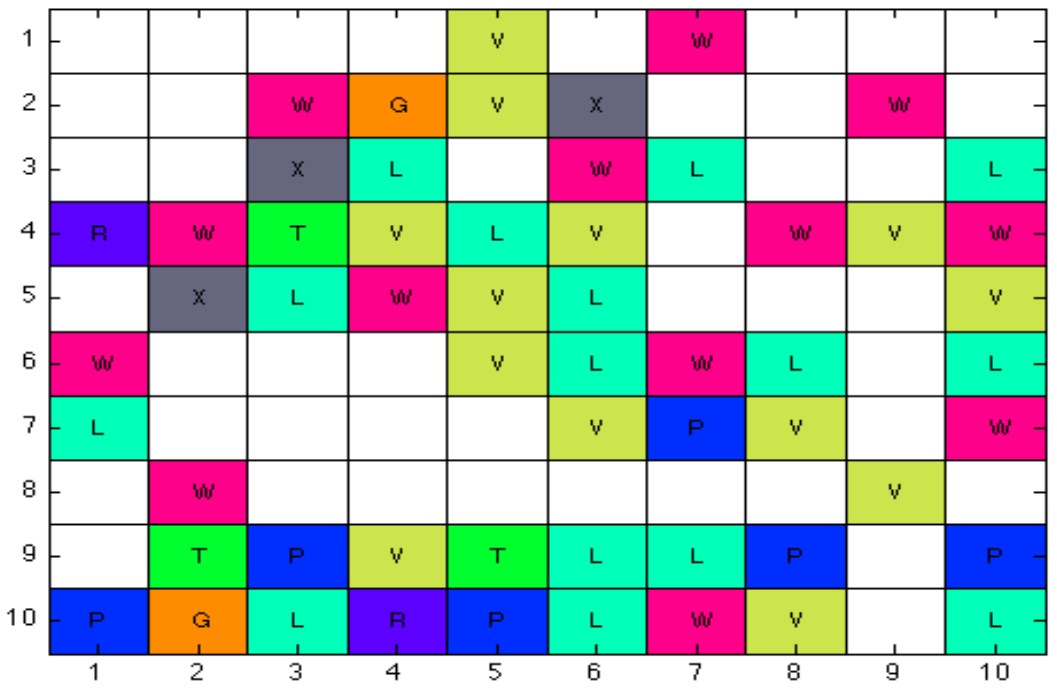

User-location behavioral groups. X and y axes represent locations and user clusters respectively

Fig. 4. Best traffic fit for user-domain and user-location behavioral groups. (L: Log-normal, G: Gamma, P: Poisson, V: Generalized Extreme Value, T: Generalized Pareto, W: Weibull, Empty cells show no traffic).

## V. ACCURACY ANALYSIS

In this section, we compare the accuracy of generic modeling vs interest-based modeling of big mobile networks. For this purpose, we calculate a weighted average of KS distances between the estimated and the actual distributions for different domains, locations and behavioral groups based on their traffic density. Table 1 shows the result of the evaluation. As can be seen, KS distance for the interest-based approach based on accessed domains, visited locations, user-domain groups and user-location groups is significantly reduced comparing to the generic approach. The analysis results show that if we use the generic model to reproduce traffic distribution of different domains, locations or user behavioral groups, the KS distance will be significantly large. As can be seen, the average KS distance for domain and location based modeling is more than 56 and 74 percent respectively. This measure for behavioral groups is more than 77 percent for user-domain groups and close to 62 percent for user-location groups. However, if we use the proposed interest-based modeling technique the KS distance is reduced to around 11 percent for domains, 14 percent for locations, 15 percent for user-domain groups and only 8 percent for user-location groups. This means a significant improvement by factor of 7 for user-location behavioral groups and by factor of 5 for the rest. This clearly shows the importance of interest-based mining and modeling for big mobile networks.

Table 1- Comparison of generic vs interest-based traffic modeling of big mobile networks based on KS test.

| Approach | Domain Specific | Location-based | User-Domain Groups | User-Location Groups |
|---|---|---|---|---|
| Generic | 0.5643 | 0.7427 | 0.7765 | 0.6187 |
| Interest-based | 0.1159 | 0.1402 | 0.1565 | 0.0811 |

As can be seen in Table 1, traffic modeling based on user-location behavioral groups shows the best accuracy (with the smallest KS distance of 0.0811). This shows that considering user behavioral groups in addition to location groups has a significant impact on the accuracy of traffic model comparing to the location-based modeling. However, in case of user-domain groups, we do not see an improvement comparing to the domain-based approach. One reason for this may be the fact we observed in the previous section that the traffic distribution of different user-domain groups are more driven by the domains rather than users. However, for user-location groups the behavior of users plays an important role in the traffic characteristics of multi-modal groups.

## VI. CONCLUSION

This study is motivated by the need for developing realistic mining and modeling techniques for big mobile networks. For this purpose, we proposed an interest-based approach based on accessed domains, visited locations and user behavioral groups. Using a novel graph-based approach and a co-clustering technique we showed that the characteristic of big mobile network traffic largely depends on users' behavior and interests captured from domain accesses and location visitations. We also showed that the proposed interest-based approach can significantly improve the modeling accuracy of big mobile networks which is essential to the design of future mobile services and protocols.